\documentclass[%
reprint,
amsmath,amssymb,
aps,
prd,
nofootinbib, 
]{revtex4-2}

\usepackage{graphicx}
\usepackage{dcolumn}
\usepackage{bm}
\usepackage{color}
\usepackage{lipsum}
\usepackage{hyperref}

\usepackage{array, multirow} 
\usepackage{hhline}

\usepackage{relsize,exscale}
\usepackage{array}
\usepackage{ctable}
\usepackage{cellspace} 
\newcolumntype{?}{!{\vrule width 0.12em}}

\newcommand{\class}{\textsc{class}}
\newcommand{\nede}{_{_{\rm NEDE}}}

\usepackage[normalem]{ulem}

\begin{document}

\title{Profiling Cold New Early Dark Energy}

\author{Juan S. Cruz$^1$}
\email{jcr@sdu.dk}
\author{Steen Hannestad$^2$}
\email{sth@phys.au.dk}
\author{Emil Brinch Holm$^2$}
\email{ebholm@phys.au.dk (corresponding author)}
\author{Florian Niedermann$^3$}
\email{florian.niedermann@su.se}
\author{Martin S. Sloth$^1$}
\email{sloth@cp3.sdu.dk}
\author{Thomas Tram$^2$}
\email{thomas.tram@phys.au.dk}

\affiliation{
	\vspace{0.2cm}
	$^1$CP$3$-Origins, Center for Cosmology and Particle Physics Phenomenology, University of Southern Denmark, Campusvej 55, 5230 Odense M, Denmark
}
\affiliation{
	\vspace{0.2cm}
	$^2$Department of Physics and Astronomy, Aarhus University, DK-8000 Aarhus C, Denmark
}
\affiliation{
	\vspace{0.2cm}
	$^3$Nordita, KTH Royal Institute of Technology and Stockholm University, \\Hannes Alfvéns väg 12, SE-106 91 Stockholm, Sweden
}

\date{\today}

\begin{abstract}
	Recent interest in New Early Dark Energy (NEDE), a cosmological model with a vacuum energy component decaying in a triggered phase transition around recombination, has been sparked by its impact on the Hubble tension. Previous constraints on the model parameters were derived in a Bayesian framework with Markov-chain Monte Carlo (MCMC) methods. In this work, we instead perform a frequentist analysis using the profile likelihood in order to assess the impact of prior volume effects on the constraints. We constrain the maximal fraction of NEDE $f_\mathrm{NEDE}$, finding $f_\mathrm{NEDE}=0.076^{+0.040}_{-0.035}$ at $68 \%$ CL with our baseline dataset and similar constraints using either data from SPT-3G, ACT or full-shape large-scale structure, showing a preference over $\Lambda$CDM even in the absence of a SH0ES prior on $H_0$. While this is stronger evidence for NEDE than obtained with the corresponding Bayesian analysis, our constraints broadly match those obtained by fixing the NEDE trigger mass. Including the SH0ES prior on $H_0$, we obtain $f_\mathrm{NEDE}=0.136^{+0.024}_{-0.026}$ at $68 \%$ CL. Furthermore, we compare NEDE with the Early Dark Energy (EDE) model, finding similar constraints on the maximal energy density fractions and $H_0$ in the two models. At $68 \%$ CL in the NEDE model, we find $H_0 = 69.56^{+1.16}_{-1.29} \text{ km s}^{-1}\text{ Mpc}^{-1}$  with our baseline and $H_0 = 71.62^{+0.78}_{-0.76} \text{ km s}^{-1}\text{ Mpc}^{-1}$ when including the SH0ES measurement of $H_0$, thus corroborating previous conclusions that the NEDE model provides a considerable alleviation of the $H_0$ tension. 
\end{abstract}

\maketitle

\section{Introduction}\label{sec:level1}
The well-known 5-$\sigma$ discrepancy between the SH0ES collaboration's measurement of the Hubble constant of~\cite{Riess:2021jrx} $H_0 = 73.04 \pm 1.04$ km s$^{-1}$ Mpc$^{-1}$ using type Ia supernovae (SNe) up to a redshift of order one and the similar measurement of the Hubble constant by the Planck collaboration using the cosmic microwave background (CMB) of~\cite{Planck:2018vyg} $H_0 = 67.36 \pm 0.54$ km s$^{-1}$ Mpc$^{-1}$ is model dependent. That is because the latter measurement assumes the $\Lambda$CDM model in order to propagate the CMB signal from the time of last scattering until today. There is, therefore, hope that a modification of $\Lambda$CDM can resolve the Hubble tension (for reviews see Refs.~\cite{Knox:2019rjx,DiValentino:2020zio,Schoneberg:2021qvd,Abdalla:2022yfr}).

However, a different, rather model-independent, constraint severely limits the type of modifications of $\Lambda$CDM, which can successfully resolve the tension. From fits to Baryonic Acoustic Oscillations (BAO) and Pantheon SNe we learn that there is a degeneracy between $H_0$ and the sound horizon $r_s$, which requires $H_0\propto 1/r_s$~\cite{Bernal:2016gxb,Aylor:2018drw,Knox:2019rjx,Efstathiou:2021ocp}. Thus any model that fits the CMB with a larger value of $H_0$ needs to do so with a smaller value of the sound horizon in order to satisfy this constraint, indicating that new physics before recombination is required in order to accommodate the tension. New late-time physics after recombination, such as phantom dark energy or late-time modified gravity models~\cite{Bernal:2016gxb,Aylor:2018drw,Knox:2019rjx,Efstathiou:2021ocp}, are therefore severely obstructed as a solution to the Hubble tension.

Several proposals for new pre-recombination physics have been put forward as solutions to the Hubble tension. But simple proposals, such as adding an extra component of non-interacting dark radiation, have the problem that their effect on the CMB spectrum are too large and, therefore, they are over-constrained as a solution to the Hubble tension~\cite{Planck:2018vyg}, although interacting scenarios provide more freedom~\cite{Brust:2017nmv,Blinov:2020hmc,Corona:2021qxl,Berryman:2022hds}. One class of models capable of modifying the expansion history sufficiently to solve the Hubble tension while only leading to a small localized effect on the CMB has nevertheless emerged as particularly successful. In this class of models, one has a new component of Dark Energy, which decays just before it gets to dominate the energy density of the universe slightly before recombination.

In the first generation of this type of model~\cite{Karwal:2016vyq, Poulin:2018cxd,Poulin:2018dzj,Smith:2019ihp,Kamionkowski:2022pkx}, the EDE-model, the new dark energy component decays as a scalar field, which is initially frozen in its potential due to Hubble friction, becomes heavier than the Hubble rate and slowly rolls towards the minimum of its potential where it starts to oscillate. For the decay of the scalar field, and the effect on the CMB to be sufficiently localized to solve the Hubble tension, this class of models requires some fine-tuning of the initial condition of the field and the shape of the potential around the minimum~\cite{Kaloper:2019lpl}. This motivated another group to introduce a more natural model of early-type dark energy, the NEDE-model~\cite{Niedermann:2019olb, Niedermann:2020dwg}, where the NEDE component decays in a fast-triggered phase transition (see also Refs.~\cite{Allali:2021azp, Freese:2021rjq} for closely related models, and Refs.~\cite{Niedermann:2021vgd, Niedermann:2021vgd} for further microscopic explorations of the NEDE framework). At the background level, the EDE and NEDE models have many similarities, but they are very different at the level of perturbations and imprint themselves differently on CMB and large scale structure (LSS) perturbations~\cite{Niedermann:2020qbw,Cruz:2022oqk}.

Early parameter constraints on the (N)EDE models were constructed within the Bayesian statistical paradigm, and were thus impacted by prior volume effects~\cite{Poulin:2018cxd, Niedermann:2020dwg, Herold:2021ksg, Gomez-Valent:2022hkb, Herold:2022iib}. The latter refer to the increased preference for regions in parameter space associated with large volumes of non-negligible likelihood, which are emphasised upon marginalisation. In the (N)EDE models, this occurs in the limit where the maximal energy density fraction of (N)EDE, $f_\mathrm{(N)EDE}$, vanishes, since the other model parameters, such as the decay redshift $z_\mathrm{decay}$, thus become unconstrained, enlarging the volume in parameter space that is probed by a sampling algorithm such as an MCMC. It was argued that this leads to non-Gaussian artefacts in the posterior distributions for $f_\mathrm{(N)EDE}$ and $H_0$, impeding a clear assessment of the models' ability to resolve the $H_0$ tension without using a late-time prior on $H_0$~\cite{Poulin:2018cxd,Niedermann:2020dwg}. Most notably, this issue has led to diverging claims about the models' ability to address the Hubble tension when the parameter inference includes full-shape LSS data (see the claims in Refs.~\cite{DAmico:2020ods,Hill:2020osr,Ivanov:2020ril} as opposed to Refs.~\cite{Poulin:2018dzj, Niedermann:2020dwg,Niedermann:2020qbw,Murgia:2020ryi,Smith:2020rxx}). A simple albeit ad-hoc way of dealing with this issue was proposed in Ref.~\cite{Niedermann:2020dwg} in the context of NEDE (but also used for EDE in Ref.~\cite{Murgia:2020ryi}). It consists in fixing all the NEDE parameters except $f_\mathrm{NEDE}$ close to their best-fit value. This keeps the sampling volume finite in the limit $f_\mathrm{(N)EDE} \to 0$ and hence avoids the issue. However, it comes at the price of giving up information about the covariances between (N)EDE parameters, e.g.\ between $z_\mathrm{decay}$ and  $f_\mathrm{(N)EDE}$, and reduces the generality of the model given that the bestfit values of the fixed parameters may as well vary across parameter space.

Alternatively, the sampling volume issue can be avoided by performing a frequentist analysis based on profile likelihoods~\cite{pawitan}. Here, the idea is to infer the likelihood  $L(\theta_i)$ of a given parameter $\theta_i$ by fixing all other parameters $\theta_{j\neq i}$ to their maximum likelihood estimates. This approach also avoids the limitations of the Bayesian analysis with fixed parameters and has recently been used to confirm the EDE model~\cite{Herold:2021ksg,Herold:2022iib} as a phenomenologically viable solution to the Hubble tension. The aim of this paper is to use the profile likelihood analysis to constrain the NEDE model using recent CMB, BAO, SNe, Big Bang Nucleosynthesis (BBN) and LSS data. Particular emphasis is put on establishing the effect of including full-shape LSS data and supplementing Planck data with ground-based CMB data from the Atacama Cosmology Telescope (ACT) and the South Pole Telescope (SPT). This complements and tests two recent, purely Bayesian, analyses in  Refs.~\cite{Cruz:2022oqk, Niedermann:2020qbw}. In doing so, we show that the one-dimensional marginalized posterior and profile likelihood in $f_\mathrm{NEDE}$ coincide when the trigger field mass is fixed, in support of the method described above. Moreover, we provide a direct comparison between EDE and NEDE and assess the models' differences.

In Sec.~\ref{sec:level2}, we review the phenomenological NEDE model and highlight its differences from EDE, manifesting themselves on the perturbation level.  In Sec.~\ref{sec:level3}, we introduce the profile likelihood and explain how it can be constructed using an efficient optimization procedure. We present our main results in Sec.~\ref{sec:level4}. This includes the profile likelihoods for the maximal fraction of NEDE, $f_\mathrm{NEDE}$, in Fig.~\ref{fig:intervals_profile} and $H_0$ in Fig.~\ref{fig:profile_H0}. Results are summarized in Table~\ref{summary_table}. We conclude in Sec.~\ref{sec:level5}.

\section{New early dark energy}\label{sec:level2}
The NEDE model, introduced in Refs.~\cite{Niedermann:2019olb, Niedermann:2020dwg}, falls in the category of early time modifications of $\Lambda$CDM. It suggests a solution to the Hubble tension by means of reducing the size of the sound horizon, $r_s$. On a purely phenomenological level\textemdash we discuss a microscopic model later\textemdash  the model adds a new energy component, $\rho\nede$, to $\Lambda$CDM. Initially, it behaves as dark energy up to a certain time, $t_*$, alternatively a  redshift $z_{\rm decay}$, at which it begins to redshift away.
In order to have a noticeable impact on the Hubble parameter, it is required that the decay of this new component must occur not too long before recombination,  around matter-radiation equality. Thereafter, the energy fraction stored in it starts to decay rapidly, i.e. faster than radiation; in this way, the model avoids creating big deviations in other cosmological parameters.

The equation of state of the NEDE component can be stated as
\begin{equation}
	w\nede = \begin{cases}
		-1 \hspace{1.35cm}\text{for}\quad  t < \bar{t}_* \\
		w\nede(t) \quad \text{for}\quad t\geq \bar{t}_*
	\end{cases},
\end{equation}
where $\bar{t}_*$ corresponds to the background quantity when decomposing the trigger time as $t_*({\bf x}) = \bar{t}_* + \delta t_*({\bf x})$. Here the spatial dependence of the perturbation $\delta t_*({\bf x})$ encodes the details of how the transition is triggered. It affects the decay of NEDE as seen by integrating its continuity equation (valid close to the transition surface)~\cite{Cruz:2022oqk}
\begin{equation}\label{eq:rho_NEDE}
	\rho\nede(t,{\bf x}) \simeq \bar{\rho}^*\nede \exp\left(-3\int_{t_*({\bf x})}^t \!\!\!{\rm d}\tilde{t} H(\tilde{t})(1+w\nede(\tilde{t})) \right)\,.
\end{equation}
This setup has been shown to alleviate the Hubble tension when the energy density fraction of NEDE at the time of decay is $f_{_{\rm NEDE}}\equiv \rho_{_{\rm NEDE} } (\bar{t}_*)/\rho_{\rm tot}(\bar{t}_*) \approx 10\%$~\cite{Niedermann:2020dwg}. The main features distinguishing NEDE from the earlier EDE model~\cite{Karwal:2016vyq, Poulin:2018cxd,Poulin:2018dzj} is the way the transition to the decaying stage happens. In EDE it occurs when the Hubble drag of an ultralight scalar field gets released. This means that EDE does not admit (at least initially) the above fluid description; instead, the oscillations of the scalar field need to be tracked explicitly. NEDE, on the other hand, relies on an external trigger to initiate a \textit{phase transition}, which is subsequently described as a decaying perfect fluid. Independent of the specific implementation of the NEDE trigger (cold~\cite{Niedermann:2020dwg}, hot~ \cite{Niedermann:2021ijp, Niedermann:2021vgd}, or hybrid~\cite{Niedermann:2020dwg}) and the nature of the transition (first or second order), NEDE can be described as a decaying perfect fluid after the phase transition, as in \eqref{eq:rho_NEDE}. In particular, on a field theoretic level, the model can be implemented in terms of a more natural potential.\\

The trigger is described in terms of an auxiliary field $q=q(t,{\bf x})$. It carries along adiabatic perturbations that introduce a spatial dependence, meaning different regions transition at different cosmological times. In cold NEDE, a physical, sub-dominant scalar field plays the role of the trigger, while in hot NEDE it is the temperature of a dark sector thermal bath. In all cases it makes the phenomenological model in \eqref{eq:rho_NEDE} sensitive to the underlying microscopic theory.

In the following, we briefly review the cold NEDE implementation, which is constrained in this article. In such a version of NEDE, a pair of scalar fields are responsible for achieving the behavior described above. A first scalar field, $\psi$, with a mass $M\sim$ eV, is associated with the NEDE energy density, and a second scalar field, $\phi$, with $m\sim 10^{-27}$ eV, acts as a trigger. By means of a coupled, two-field potential,
\begin{equation}
	V(\psi,\phi) = \frac{\lambda}{4}\psi^4 + \frac{1}{2}M^2\psi^2 - \frac{1}{3}\alpha M \psi^3 + \frac{1}{2}m^2\phi^2 + \frac{1}{2}\tilde{\lambda}\phi^2\psi^2\, ,
\end{equation}
the rolling down of $\phi$, which starts when $H\lesssim m$, opens up a new lower minimum for $\psi$ to tunnel into and permits the first-order phase transition to take place. The phase transition proceeds in the usual sense, by the nucleation of expanding bubbles of the new phase.

Under specific requirements~\cite{Niedermann:2020dwg}, which can be cast as conditions on the dimensionless parameters $\lambda$, $\alpha$, and $\tilde \lambda$, the transition occurs very rapidly and only relatively small bubbles percolate, later covering the whole region of the universe where the transition has been triggered. The bubble collisions, interactions and evolution can be understood by means of an effective fluid description. On small scales, we expect this field condensate to be dominated by anisotropic stress, which manifests itself on large scales as a fluid that decays faster than radiation, i.e. $1/3 < w\nede(t) < 1$ with $t>\bar{t}_*$, while also producing gravitational waves and other microscopic decay products.

Parametrizing the condensate as a perturbed perfect fluid, $ \rho_\mathrm{NEDE}(t,\mathbf{x}) =\bar{\rho}_\mathrm{NEDE} + \delta\rho_\mathrm{NEDE}(t,\mathbf{x})$, one can employ perturbation matching to initialize the perturbations of the fluid with the perturbations of the trigger field. That is, by tracking the evolution of the trigger field $\phi$ and its adiabatic perturbations and employing Israel matching conditions~\cite{Israel1966}, the NEDE density fluctuation, $\delta\nede \equiv \delta\rho\nede/\bar{\rho}\nede$, and its velocity divergence $\theta\nede$, can be initialized via:
\begin{subequations}\label{eq:matching}
	\begin{align}
		\delta^*\nede &= -3 \left[1 + w\nede(t_*) \right] H_* \frac{\delta q_*}{\dot{\bar{q}}_*}\,,\\
		\theta\nede^* &= \frac{k^2}{a_*} \frac{\delta q_*}{\dot{\bar{q}}_*}\,.
	\end{align}
\end{subequations}
where the role of the trigger, $q$ is played by $\phi$, i.e. $q\equiv \phi$, in cold NEDE, and the star denotes quantities evaluated at $\bar{t}_*$. The subsequent evolution can then be carried out with the usual equations governing the dynamics of fluid perturbations~\cite{Ma:1995ey}. For the concrete case of cold NEDE, we assume vanishing viscosity and anisotropic stress (for the scales that may impact CMB), and an effective sound speed in the fluid's rest frame that equals the adiabatic sound speed,
\begin{equation}
	c_a^2 = w\nede(t) - \frac{1}{3} \frac{\dot{w}\nede(t)}{1+w\nede(t)}\frac{1}{H}\,.
\end{equation}
Finally, we assume that the NEDE equation of state is constant after the decay. At this stage, the cold NEDE setup can be implemented in a Boltzmann code to perform comparisons and analysis against different data sources (for more details on the implementation and microscopic picture of cold NEDE, see Ref.~\cite{Niedermann:2020dwg}).

\section{Profile likelihood}\label{sec:level3}
A profile likelihood $L(\theta_i)$ (PL) of a parameter $\theta_i$ is obtained from the likelihood function $L(\theta_1, ..., \theta_N)$ by fixing all parameters $\theta_j, j\neq i$ to their maximum likelihood estimate,
\begin{align} \label{eq:prof_defn}
	L(\theta_i) = \max_{\theta_j, j\neq i} L(\theta_1, \theta_2, ..., \theta_N).
\end{align}
Since for each fixed $\theta_j$, $L(\theta_j)$ is a maximum likelihood estimate, the profile likelihood inherits the reparameterization invariance of the maximum likelihood estimator~\cite{pawitan}. This property, along with the inherent independency of prior distributions, is the key difference between frequentist likelihood-based inference and Bayesian inference.

The statistical significance of the profile likelihood arises from Wilks' theorem~\cite{pawitan}, which states that the distribution of the quantity $-2 \log(L(\theta_j)/L_\text{max})$, with $L_\text{max} = \max_{\theta_j} L(\theta_j)$, asymptotes toward a $\chi^2$ distribution with one degree of freedom. We therefore write $\Delta \chi^2 (\theta_j) \equiv -2 \log(L(\theta_j)/L_\text{max})$ in the following. From this, an approximate $68 \%$ ($95 \%$) confidence interval in $\theta_j$ can be obtained as the region $\Delta \chi^2 (\theta_j) < 1.0 \ (3.84)$ according to the Neyman construction~\cite{Neyman:1937uhy}. These confidence levels are exact in the case that the profile likelihood is Gaussian, however, since the profile likelihood is reparameterization invariant, the confidence levels also hold whenever there exists a reparameterization in which the profile likelihood is Gaussian. In practice it is difficult to determine to what extent this holds, so in the following, we cite these confidence levels but acknowledge that they may only approximate the true confidence levels.

The Feldman-Cousins prescription~\cite{Feldman:1997qc} provides a confidence interval construction with more accurate coverage close to physical boundaries of the quantity under study. Since the profiles we find for the maximum fraction of NEDE, $f_\mathrm{NEDE}$ have a non-negligible intersection with the natural boundary at $f_\mathrm{NEDE} = 0$, we have computed Feldman-Cousins intervals for these profiles in a manner similar to Refs.~\cite{Herold:2021ksg, Herold:2022iib}, but find that they coincide with the Neyman intervals at $68 \%$ CL and differ only marginally at $95 \%$ CL. All results stated in this paper are therefore based on Neyman intervals.

In practice, constructing profile likelihoods consists of optimizations in an often high-dimensional, noisy likelihood function. We carry out the optimization with \textit{simulated annealing}~\cite{Kirkpatrick:1983zz}, using the same procedure as Ref.~\cite{Holm:2022kkd}. We modify the likelihood according to $L \rightarrow L^{1/T}$, where $T$ is referred to as the likelihood temperature. For $T>1$, the likelihood surface is smoothened. For $T<1$, peak structures are enhanced. Simulated annealing consists in running MCMC chains with iteratively decreased temperatures. This works since the MCMC chain is increasingly localized around the likelihood peaks as the temperature is decreased; however, due to the randomness of the MCMC chain, the algorithm may still escape local optima. Thus, simulated annealing generally works well in likelihood functions with many local optima, at the cost of relying somewhat strongly on the particular temperature values chosen. In practice, we start from the proposal covariance matrices from the Bayesian analyses of NEDE~\cite{Niedermann:2020dwg,Niedermann:2020qbw,Cruz:2022oqk} to inform the MCMC proposal distribution, and we furthermore decrease the step size in addition to the temperature in order to increase the resolution around the successively narrower likelihood peaks. In agreement with earlier results~\cite{Hannestad:2000wx, Holm:2022kkd} we found that exponentially decreasing temperatures and step sizes performed well; the particular schedule used depended on the quality of the covariance matrix employed. Our code, identical that of Ref.~\cite{Holm:2022kkd}\footnote{\href{https://github.com/AarhusCosmology/montepython\_public/tree/2211.01935}{github.com/AarhusCosmology/montepython\_public/tree/2211.01935}.}, uses \textsc{MontePython}~\cite{Audren:2012wb, Brinckmann:2018cvx} and the \class{}~\cite{Blas:2011rf} implementation of NEDE in \textsc{TriggerClass}\footnote{\href{https://github.com/flo1984/TriggerCLASS/tree/NewEDEv5.0}{github.com/flo1984/TriggerCLASS/tree/NewEDEv5.0}.}~\cite{Niedermann:2020dwg}. For the EDE-model computations, we use a slightly modified version of the \textsc{class\_ede} implementation of EDE\footnote{In shooting algorithm of the unmodified code, the EDE critical redshift $z_c$, corresponding to the maximum of the EDE energy density fraction, could only assume the discrete set of values corresponding to the ones in the internal table of \textsc{class}. We allowed $z_c$ to vary continuously by using Hermite interpolation around the maximum value. The modification can be found at \href{https://github.com/AarhusCosmology/class_ede/tree/make-shooting-continuous-and-fast}{github.com/AarhusCosmology/class\_ede/tree/make-shooting-continuous-and-fast}. For an alternative implementation of EDE, see~\cite{Poulin:2018dzj}.}~\cite{Hill:2020osr}.

\section{Results}\label{sec:level4}
In this section, we present the results of a series of profile likelihoods in different parameters and using different datasets. In the general case, we vary all of the usual cosmological parameters
\begin{align}
	\{ \omega_b, \omega_\mathrm{cdm}, H_0, \ln 10^{10} A_s, n_s, \tau_\mathrm{reio} \}.
\end{align}
In addition, we vary all the nuisance parameters required for the datasets we employ. The NEDE sector is parameterized in terms of the maximal energy density fraction of NEDE, $f_\mathrm{NEDE}$, the logarithm of the redshift of the onset of the decay, $\log_{10} z_\mathrm{decay}$, and the NEDE equation of state, $w_\mathrm{NEDE}$. Except for those fixed in a particular profile likelihood, all of these are varied in our analysis.

Our baseline dataset consists of the following:
\begin{itemize}
	\item \textit{Planck} 2018 high-$\ell$ TTTEEE, low-$\ell$ TT and EE and lensing data~\cite{Planck:2018vyg}.

	\item BAO data, including BOSS DR12~\cite{BOSS:2016wmc} and low redshift data from 6dF~\cite{Beutler:2011hx} and the BOSS main galaxy sample~\cite{Ross:2014qpa}, as well as growth structure measurements from the CMASS and LOWZ galaxy samples of BOSS DR12~\cite{BOSS:2016wmc}.

	\item The Pantheon catalogue of type Ia supernovae in the redshift range $0.01 < z < 2.3$~\cite{Pan-STARRS1:2017jku}.

	\item A Gaussian likelihood on the primordial helium abundance $Y_p = 0.2449 \pm 0.0040$ from the measurements of reference~\cite{Aver:2015iza}.
\end{itemize}
We note that this baseline coincides with the baseline of Ref.~\cite{Niedermann:2020dwg} up to the particular prior used on $Y_p$, but we expect this difference to contribute negligibly to the constraints obtained. In addition to the baseline, we employ the following datasets when specified.
\begin{itemize}
	\item \textbf{SH0ES}: A Gaussian likelihood on the value of $H_0 = 73.04\pm 1.04$ km s$^{-1}$ Mpc$^{-1}$ as measured by the SH0ES collaboration~\cite{Riess:2021jrx}. We note that putting the likelihood on $H_0$ directly instead of the calibration of the intrinsic SNIa magnitude $M_b$ nuisance parameter of the Pantheon dataset is appropriate since in all parameter space regions of interest, the NEDE model does not radically alter the luminosity distance at small redshifts~\cite{Benevento:2020fev, Camarena:2021jlr}.

	\item \textbf{FS}: Full-shape analysis of the LOWZ and CMASS redshift splits of the monopole and quadrupole of the BOSS power spectrum~\cite{Zhang:2021yna} with reconstruction of the BAO peak scale from Ref.~\cite{Gil-Marin:2015nqa}. This includes a consistent normalization of the window function following the prescription of Ref.~\cite{Beutler:2021eqq}, contrary to the previous NEDE result of Ref.~\cite{Niedermann:2020qbw}. We have taken a scale cut of $k_\mathrm{max}=0.25h$ Mpc$^{-1}$ similarly to previous studies of EDE~\cite{Herold:2022iib, Simon:2022adh, Herold:2021ksg}. The weakly non-linear theory power spectrum out to this scale cut is computed from the effective field theory of large-scale structure (EFTofLSS) by the \textsc{PyBird} code~\cite{DAmico:2020kxu}\footnote{\href{https://github.com/pierrexyz/pybird}{github.com/pierrexyz/pybird}}. We use the standard \textsc{PyBird} EFT parameterization (the \textit{West-coast} parameterization) as well as the standard \textsc{PyBird} priors on the EFT parameters with analytical marginalisation over those with Gaussian priors~\cite{DAmico:2020kxu, DAmico:2021ymi}\footnote{Although marginalisation is ill-defined in the frequentist approach, it increases computational efficiency and should not have a large impact on the results~\cite{DAmico:2020kxu}}. Although these choices may generally affect final results~\cite{Simon:2022lde, DAmico:2022osl}, they should have little impact in our analysis since we always pair the FS data with the strongly constraining baseline dataset. Finally, when using this FS data, we omit the BAO and growth structure data in the baseline to avoid double counting.

	\item \textbf{ACT}: Temperature and polarization anisotropy measurements of the CMB spectrum by the Atacama Cosmology Telescope~\cite{ACT:2020frw, ACT:2020gnv}. In order to avoid double counting of the Planck-range multipoles, we exclude the ACT multipoles $\ell \leq 1800$ as suggested by the ACT collaboration. We employ the \textsc{pyactlike} likelihood provided by the ACT collaboration\footnote{\href{https://github.com/ACTCollaboration/pyactlike}{github.com/ACTCollaboration/pyactlike}}. The ACT collaboration suggests using a strict set of \class{} precision parameters for the EDE-model~\cite{Hill:2021yec}; however, we find only little variation in the $\chi^2$ values obtained with the default settings and therefore use those in the interest of computational resources.

	\item \textbf{SPT}: The year one data release of the South Pole Telescope EE and TE power spectra at multipoles $300 \leq \ell < 3000$~\cite{SPT-3G:2014dbx}. We use the \textsc{clik} implementation provided by the SPT collaboration\footnote{\href{https://github.com/SouthPoleTelescope/spt3g_y1_dist}{github.com/SouthPoleTelescope/spt3g\_y1\_dist}}.
\end{itemize}
We take wide, uniform priors on all of the cosmological parameters and include two massless and one massive (with $m=0.06$ eV) neutrino species equivalent to the $N_\mathrm{eff}$ measurement of the \textit{Planck} collaboration~\cite{Planck:2018vyg}. Since Ref.~\cite{Reeves:2022aoi} found this to coincide with the bestfit configuration, we do not view it as a restriction of the cosmological model used. Table~\ref{summary_table} summarizes all constraints obtained in this paper.

\subsection{Constraints on the maximal fraction of NEDE}
\begin{figure}[tb]
	\includegraphics[width=\columnwidth]{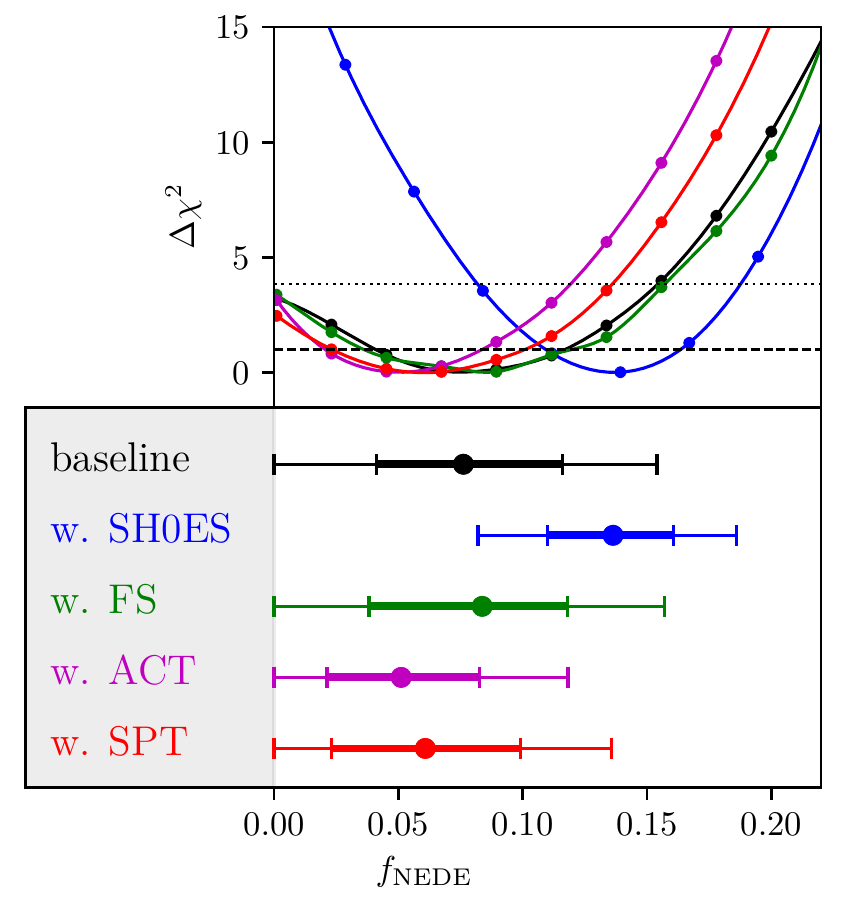}
	\caption{\label{fig:intervals_profile} \textit{Top panel}: Profile likelihoods for five different data combinations as specified by the legend in the bottom panel. The horizontal dashed and dotted lines represent the lines $\Delta \chi^2 = 1.0$ and $\Delta \chi^2 = 3.84$, giving the approximate $68 \%$ and $95 \%$ confidence intervals, respectively, from the intersection with the profiles. \textit{Bottom panel}: Confidence intervals for the maximum fraction of NEDE, $f_\mathrm{NEDE}$, for different combinations of datasets, computed with the Neyman construction from profile likelihoods. The points correspond to best-fit values and the inner (outer) error bars represent (approximate) $68\%$ ($95\%$) confidence levels.}
\end{figure}
Firstly, we present profile likelihood constraints on the maximum fraction of NEDE, $f_\mathrm{NEDE}$. The top panel in Fig.~\ref{fig:intervals_profile} illustrates profile likelihoods for the baseline (black), baseline including SH0ES (blue), baseline including FS (green), baseline including ACT (magenta) and baseline including SPT (red). The points correspond to the fixed values of $f_\mathrm{NEDE}$ at which we have computed the profile, and the fully drawn lines represent cubic interpolations between them. The horizontal dashed and dotted lines correspond to $\Delta \chi^2 = 1.0$ and $\Delta \chi^2 = 3.84$, such that the intersection between these and the profiles define the approximate $68 \%$ and $95 \%$ confidence intervals, respectively. The intervals obtained from each profile is shown in the bottom panel for each data combination, with the inner (outer) error bars representing the approximate $68 \%$ ($95 \%$) confidence levels and the points marking the global bestfit values of $f_\mathrm{NEDE}$. The profiles are evaluated at a set of $\approx 10$ values of $f_\mathrm{NEDE}$ and interpolated cubically. Since cubic interpolation rarely creates new minima, this results in poor bestfit resolution, so our bestfit values instead have been obtained as the minimum of an exact parabolic fit to the three points of smallest $\Delta \chi^2$ values in each profile. 

The overall picture is that the inclusion of either of FS, ACT and SPT has little impact compared to the strongly constraining baseline. Indeed, all of these find a bestfit that is non-zero, with two-sided bounds on $f_\mathrm{NEDE}$ at $68\%$ CL but upper bounds at $95\%$ CL.

Generally, these constraints allow larger values of $f_\mathrm{NEDE}$ than suggested by the Bayesian analyses of Refs.~\cite{Niedermann:2020dwg,Niedermann:2020qbw,Cruz:2022oqk}. The frequentist analysis therefore suggests a greater ability of the model to alleviate the $H_0$ tension than the corresponding Bayesian analysis. This pattern also emerges in the EDE-model~\cite{Herold:2021ksg,Herold:2022iib}, and is explained by the presence of volume effects in the NEDE sector: In the $\Lambda$CDM limit where $f_\mathrm{NEDE}$ approaches $0$, the additional model parameters $3w_\mathrm{NEDE}$ and $\log_{10} z_\mathrm{decay}$ become unconstrained and significantly increase the volume of the posterior around $\Lambda$CDM, leading to a bias toward $\Lambda$CDM when $3w_\mathrm{NEDE}$ and $\log_{10} z_\mathrm{decay}$ are marginalised over. The presence of this volume effect was already noted in Ref.~\cite{Niedermann:2020dwg}, who bypassed it by fixing one of the NEDE model parameters at its bestfit. In section~\ref{sec:baseline} we evaluate the correctness of this approach. We note that similar arguments were recently used to explain volume effects in the EDE-model~\cite{Herold:2022iib} and decaying dark matter model~\cite{Holm:2022kkd}, and that this phenomenon is to be expected in any $\Lambda$CDM extension involving additional parameters that become unconstrained in the $\Lambda$CDM limit.

Below, we discuss the result from each data combination individually.

\subsubsection{Baseline} \label{sec:baseline}
\begin{figure}[tb]
	\includegraphics[width=\columnwidth]{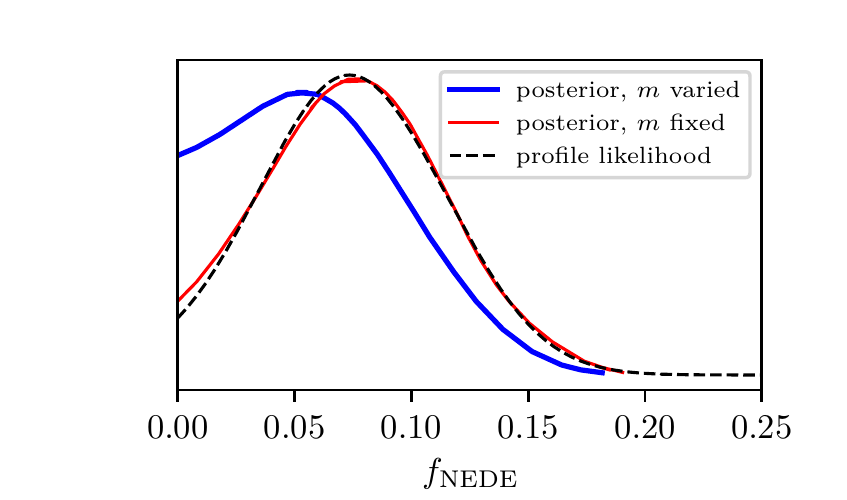}
	\caption{\label{fig:posterior_comparison} One-dimensional marginalised posteriors in $f_\mathrm{NEDE}$ from Ref.~\cite{Niedermann:2020dwg} while varying the NEDE trigger field mass $m$ (blue) and while fixing it close to its bestfit value (red), using the baseline dataset; as well as a profile likelihood in the form of eq.~\eqref{eq:prof_defn} (dashed black line). Evidently, the profile likelihood coincides with the posterior obtained when fixing the NEDE trigger field mass, showing that the latter is an appropriate method of avoiding volume effects.}
\end{figure}

With baseline data, we find the $68 \%$ confidence interval $f_\mathrm{NEDE}\in [ 0.041, 0.116]$. As noted above, our baseline coincides with the baseline of Ref.~\cite{Niedermann:2020dwg}, who found the Bayesian credible interval $f_\mathrm{NEDE}\in [0.037, 0.115] $ when fixing the value of the trigger field mass close to its bestfit value. These intervals are thus identical up to the accuracy of the simulated annealing and MCMC algorithm used to produce them. As explained in Ref.~\cite{Niedermann:2020dwg}, fixing the trigger field mass $m$ (or the decay redshift $z_\mathrm{decay}$, equivalently) removes the large posterior volume in the $f_\mathrm{NEDE}\sim 0$ region of parameter space occurring from $m$ being unconstrained in this limit. To test the validity of this work-around, we have plotted in Fig.~\ref{fig:posterior_comparison} the (arbitrarily normalized) profile likelihood $L(f_\mathrm{NEDE})$ from eq.~\eqref{eq:prof_defn} along with the Bayesian one-dimensional marginalized posteriors in $f_\mathrm{NEDE}$ from Ref.~\cite{Niedermann:2020dwg} both with and without fixing the trigger field mass. Evidently, the profile likelihood, which is inherently free from volume effects, coincides almost exactly with the marginalized posterior with fixed trigger mass. On the other hand, the posterior with a varied trigger field mass is biased toward the $f_\mathrm{NEDE}\rightarrow 0$ region due to the volume effect described. We conclude that fixing the trigger field mass, as done in Refs.~\cite{Niedermann:2020dwg, Niedermann:2020qbw, Cruz:2022oqk}, makes the Bayesian credible intervals agree with the frequentist confidence intervals.

By evaluating the profile at $f_\mathrm{NEDE} = 0$, we can estimate the bestfit $\chi^2$ value of the baseline dataset under the $\Lambda$CDM model. Hence, we get a mild statistical preference of NEDE over $\Lambda$CDM at $ \chi^2_{\mathrm{min,}\Lambda\mathrm{CDM}} - \chi_{\mathrm{min,NEDE}}^2 = -3.2$, similar to the value $-2.9$ obtained in Ref.~\cite{Niedermann:2020dwg}. Since the model extends $\Lambda$CDM with three new parameters, $\Delta \chi^2 = -3.2$ sits just below the $1\sigma$ significance level as a model nested in $\Lambda$CDM.

\subsubsection{Baseline$+$SH0ES}\label{sec:baseline_sh0es}
Adding the SH0ES likelihood increases the preference for large values of $f_\mathrm{NEDE}$ due to the well-known correlation between $f_\mathrm{NEDE}$ and $H_0$~\cite{Niedermann:2020dwg}. We believe this combined analysis is justified as the baseline constraint, given in Table~\ref{summary_table}, implies a reduced tension of approximately $2.1 \sigma$. In any event, this data combination allows the computation of the quantity~\cite{Schoneberg:2021qvd, Raveri:2018wln}
\begin{align}
	Q_\mathrm{DMAP} = \left( \chi^2_\mathrm{min,baseline+SH0ES} - \chi^2_\mathrm{min,baseline} \right)^{1/2} \nonumber
\end{align}
which quantifies the extent to which the model is able to reduce the inconsistency between the baseline and the SH0ES likelihood, and is often used to assess the ability of $\Lambda$CDM extensions to resolve the $H_0$ tension~\cite{Schoneberg:2021qvd}. By evaluating the profiles in $f_\mathrm{NEDE}$ at $f_\mathrm{NEDE}=0$, we can approximate the bestfit $\chi^2$ value for $\Lambda$CDM, and thereby compute $Q_\mathrm{DMAP}$ in the $\Lambda$CDM model as
\begin{align}
	Q_{\mathrm{DMAP,}\Lambda\mathrm{CDM}} =& \big(\Delta \chi^2_{\mathrm{baseline}} - \Delta \chi^2_{\mathrm{baseline+SH0ES}} \nonumber\\
	&+ Q_\mathrm{DMAP,NEDE}^2 \big)^{1/2} \nonumber
\end{align}
where $\Delta \chi^2 \equiv \chi^2_\mathrm{min} (\mathrm{NEDE}) -\chi^2_\mathrm{min} (\Lambda\mathrm{CDM})$ denotes the difference in bestfit $\chi^2$ values between the NEDE and $\Lambda$CDM models.
We find $Q_\mathrm{DMAP,NEDE} = 2.1$ for the NEDE model, in excellent agreement with the Gaussian tension measure, and $Q_{\mathrm{DMAP,}\Lambda\mathrm{CDM}} = 4.8$ for the $\Lambda$CDM model, corresponding approximately to a $\sim 2.7\sigma$ alleviation of the tension between the baseline and the SH0ES likelihood.

\subsubsection{Baseline$+$FS}
Using full-shape BOSS data out to $k_\mathrm{max}=0.25 h$ Mpc$^{-1}$, we obtain the confidence interval $f_\mathrm{NEDE} \in [0.038,0.118]$ at $68\%$ CL and $f_\mathrm{NEDE} < 0.157$ at $95\%$ CL. These confidence intervals are very similar to those obtained without FS, $f_\mathrm{NEDE}\in [0.041, 0.116]$ at $68 \%$ CL and $f_\mathrm{NEDE} < 0.154$ at $95\%$ CL; arguably identical within the uncertainty of the simulated annealing algorithm. This indicates that the constraining power of large scale structure is still largely dominated by the information in the BAO peak, which is also included in the baseline, and that the additional full-shape data is subdominant. A similar conclusion was reached in Ref.~\cite{Niedermann:2020qbw}.
\begin{figure}[tb]
	\includegraphics[width=\columnwidth]{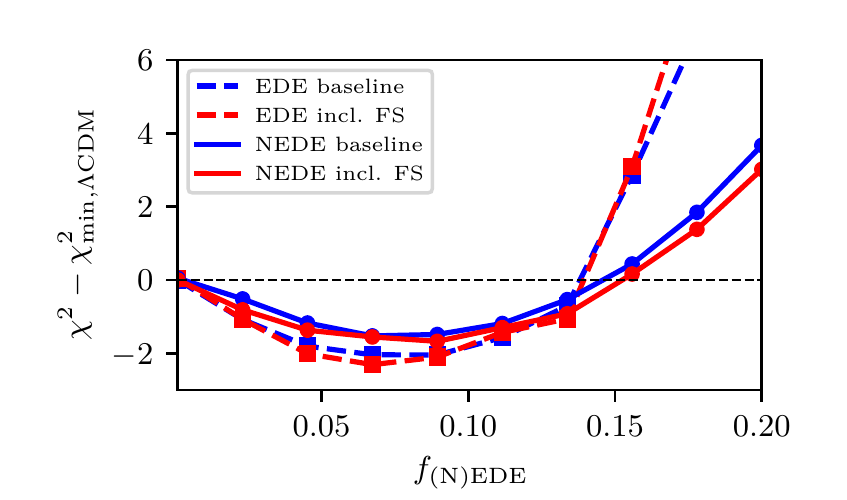}
	\caption{\label{fig:lss} Profile likelihoods for the NEDE and EDE models using the baseline (blue) and FS (red) datasets, respectively. The difference between the latter is that our FS analysis includes full-shape information on the power spectrum as predicted by EFTofLSS. There is no gain in constraining power for either model when including the full-shape analysis.}
\end{figure}
Figure~\ref{fig:lss} illustrates the profile likelihoods of $f_\mathrm{(N)EDE}$ for NEDE and EDE with and without the inclusion of full-shape data. Evidently, we do not find a statistically significant increase in the constraining power when including full-shape data. We note that the EDE constraints obtained with FS broadly match those of Ref.~\cite{Herold:2022iib}.

Ref.~\cite{Niedermann:2020qbw}, who obtained the Bayesian credible interval $f_\mathrm{NEDE}\in [0.087, 0.150]$ using EFTofLSS, found a larger preference for NEDE since they included a Gaussian prior on $H_0$ from the SH0ES measurement in order to evade volume effects. Additionally, we note that Ref.~\cite{Niedermann:2020qbw} included an inconsistently normalized window function~\cite{Simon:2022lde} (based on an earlier version of PyBird), which we have corrected. Ref.~\cite{Herold:2022iib} note that the correctly normalized data prefers a larger value of $\sigma_8$. Since $f_\mathrm{NEDE}$ is known to correlate positively with $\sigma_8$~\cite{Niedermann:2020dwg}, we expect the correction of the normalization to have increased the preference for NEDE in comparison with the inconsistent normalization. In conclusion, our result represents the first constraint on $f_\mathrm{NEDE}$ using correctly normalized EFTofLSS, without the SH0ES prior, that is free from volume effects.

\subsubsection{Baseline$+$ACT}\label{sec:fNEDE_ACT}
Including ACT data, we obtain the $68\%$ confidence interval $f_\mathrm{NEDE} \in [ 0.021, 0.083 ]$ and a weak preference of $\chi_{\mathrm{min,}\Lambda\mathrm{CDM}}^2 -  \chi^2_{\mathrm{min,NEDE}} = -3.2$. Although the level of the preference is the same as for the baseline only, the confidence interval including ACT lies at smaller values of $f_\mathrm{NEDE}$ than the one using only the baseline, which was also found in the Bayesian analysis of Ref.~\cite{Cruz:2022oqk}. There is a very precise reason for this, which we return to in section~\ref{sec:constraints_on_other_parameters}.

Our confidence interval may be compared to the Bayesian credible intervals of Ref.~\cite{Cruz:2022oqk}, derived with the same data combination. The latter find the $68 \%$ credible interval $f_\mathrm{NEDE}\in [0.011, 0.082]$ when varying all NEDE parameters. Interestingly, this interval has the same upper bound as our frequentist interval, but a smaller lower bound, corroborating the earlier sentiment that volume effects act to increase the favour for small values of $f_\mathrm{NEDE}$. When fixing the trigger field mass, Ref.~\cite{Cruz:2022oqk} found the $68\%$ credible interval $f_\mathrm{NEDE}\in [0.0336, 0.1003]$, which lies at substantially larger values than our frequentist interval. The reason is that by fixing the trigger field mass, a local optimum occurring from ACT, which favours smaller values of $f_\mathrm{NEDE}$, is excluded. Indeed, we note that the latter interval approximately coincides with our baseline confidence interval since the fixing of the trigger mass essentially excludes the ACT contribution to the likelihood. We elaborate on this point in section~\ref{sec:constraints_on_other_parameters}.

\subsubsection{Baseline$+$SPT}
With baseline and SPT data, we obtain the $68\%$ confidence interval $f_\mathrm{NEDE} \in [0.023,0.100]$ and a weak preference of $\chi_{\mathrm{min,}\Lambda\mathrm{CDM}}^2 -  \chi^2_{\mathrm{min,NEDE}} = -2.5$ for NEDE relative to $\Lambda$CDM. This interval has the same lower bound as the interval derived from the baseline$+$ACT data combination but a larger upper bound, and similarly to ACT generally lies at slightly smaller values of $f_\mathrm{NEDE}$ than the baseline itself. This is consistent with the pattern found in Ref.~\cite{Cruz:2022oqk}. The latter fixed the NEDE equation of state at $w_\mathrm{NEDE} = 2/3$, consistent with its bestfit value (see section\ref{sec:constraints_on_other_parameters}), in order to avoid the usual volume effects. In comparison, they obtain the $68\%$ Bayesian credible interval $f_\mathrm{NEDE}\in [0.014, 0.086]$ without fixing the NEDE trigger mass and $f_\mathrm{NEDE}\in [0.030, 0.101]$ while fixing it. Evidently, our frequentist results are similar to the results of Ref.~\cite{Cruz:2022oqk} with fixed NEDE trigger mass and equation of state.

\subsection{Comparison with EDE}\label{sec:ede}
Although NEDE has been compared to EDE at several occasions~\cite{Poulin:2021bjr, Schoneberg:2021qvd}, these comparisons use Bayesian inference which, as explained above, can be heavily influenced by the choice of parameters describing the cosmological model. For example, the severity of volume effects may vary strongly with different parametrizations. Thus, we present here a comparison of constraints on $f_\mathrm{(N)EDE}$ and $H_0$ between EDE and NEDE using profile likelihoods in order to circumvent any effects related to the choice of parametrizations and Bayesian priors in the models.
\begin{figure}[tb]
	\includegraphics[width=\columnwidth]{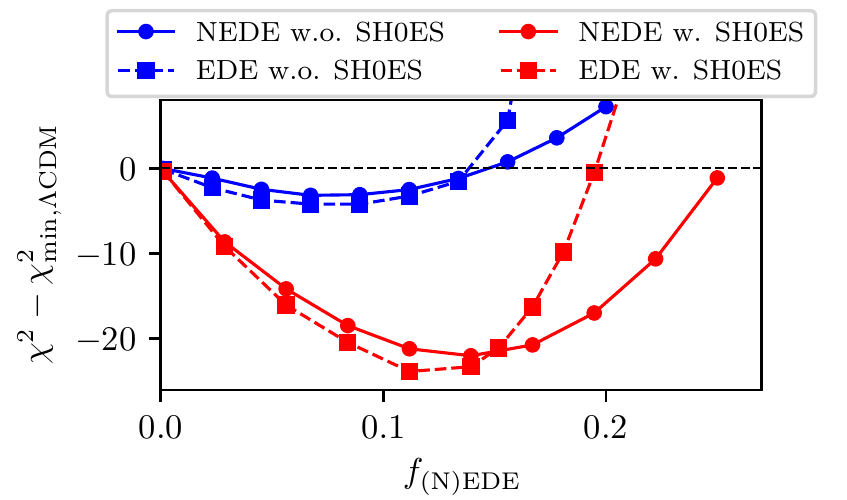}
	\caption{\label{fig:f_comparison} Profile likelihoods of the maximal energy density fraction $f_\mathrm{NEDE}$ in the NEDE model and the maximal energy density fraction $f_\mathrm{EDE}$ in the EDE model from the baseline data with and without a Gaussian likelihood on the SH0ES measurement of $H_0$. The profiles are normalized according to the $\chi^2$ value of the bestfit $\Lambda$CDM cosmology, such that they intersect at the origin.}
\end{figure}

Fig.~\ref{fig:f_comparison} shows the profile likelihoods in $f_\mathrm{EDE}$ and $f_\mathrm{NEDE}$ both without (blue) and including (red) a likelihood on the SH0ES measurement of $H_0$. The profiles are normalized differently than otherwise in this paper; the $\chi^2$ values from the optimizations are subtracted the global bestfit of the $\Lambda$CDM model. Hence, all profiles intersect at the origin. Firstly, we note that, as expected, larger values of $f_\mathrm{(N)EDE}$ are preferred when including SH0ES, and the improvement of the models over $\Lambda$CDM increases accordingly. Within each data combination, $f_\mathrm{EDE}$ and $f_\mathrm{NEDE}$ have similar constraints, with the slight systematic difference that the NEDE model admits larger values of $f_\mathrm{NEDE}$ than the values of $f_\mathrm{EDE}$ admitted by the EDE model. This may be related to the fact that data usually prefers NEDE to decay earlier than EDE~\cite{Cruz:2022oqk}, which would require NEDE to have a larger abundance in order to obtain the same energy density around recombination as EDE. However, this difference is still small, and since $f_\mathrm{NEDE}$ and $f_\mathrm{EDE}$ are in principle two different parameters in the two models, there is no reason that their constraints should coincide. Globally, EDE provides a marginally better fit to data than NEDE, with a difference in $\chi^2$ values of $\chi^2_\mathrm{min,NEDE} - \chi^2_\mathrm{min,EDE} \approx -1.0$. The numerical constraints are given in Table~\ref{summary_table}.
\begin{figure}[tb]
	\includegraphics[width=\columnwidth]{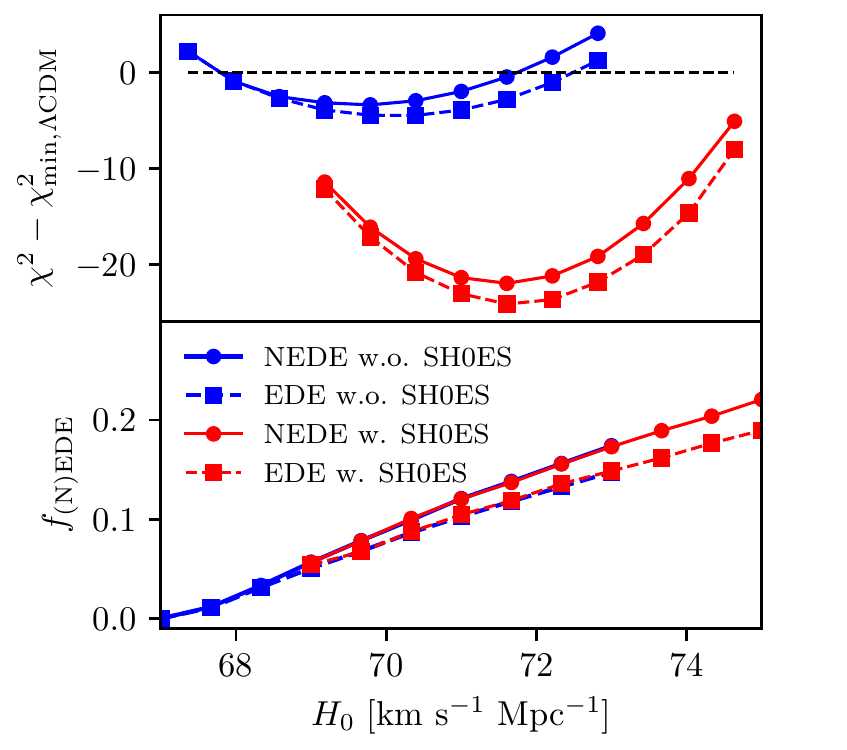}
	\caption{\label{fig:profile_H0} \textit{Top panel}: Profile likelihoods of the Hubble constant $H_0$ in the NEDE and EDE models from the baseline data with and without a likelihood on the SH0ES measurement of $H_0$, respectively. The profiles are normalized according to the $\chi^2$ value of the bestfit $\Lambda$CDM cosmology. \textit{Bottom panel}: Values of the maximal energy density fraction $f_\mathrm{(N)EDE}$ in the (N)EDE-model as obtained from optimization at each point in the profiles, illustrating the correlation between $H_0$ and $f_\mathrm{(N)EDE}$.}
\end{figure}

The difference in abundances between EDE and NEDE seen in Fig.~\ref{fig:f_comparison} is not very significant, and if it were, it could not be used to distinguish the models since the early dark energy abundance is not an observable quantity itself. Arguably, the biggest observable impact of this difference should be on the predictions for $H_0$ in the two models. Therefore, we have computed profile likelihoods of $H_0$ in the two models, respectively, with and without a Gaussian likelihood on the SH0ES measurement of $H_0$. They are shown in the top panel of Fig.~\ref{fig:profile_H0} with the same normalization as the $f_\mathrm{(N)EDE}$ profiles above. The inferences that include the SH0ES measurement naturally prefer larger values of $H_0$ leading to a larger improvement of the NEDE/EDE models over $\Lambda$CDM due to their ability to increase $H_0$. Between the NEDE and EDE models, constraints on $H_0$ are similar, with EDE constraints slightly relaxed at values of $H_0$ larger than the bestfit. The result that EDE provides a slightly better fit to data than NEDE is also clear from these profiles. The second panel in Fig.~\ref{fig:profile_H0} shows the values of $f_\mathrm{EDE}$ and $f_\mathrm{NEDE}$ obtained by optimization at each point in the profiles of the EDE and NEDE models, respectively. There is a clear one-to-one correspondence between the values of $H_0$ and $f_\mathrm{(N)EDE}$ at each fixed $H_0$, which is a manifestation of the strong correlation between the two parameters~\cite{Cruz:2022oqk,Poulin:2018cxd}. Apparently, the relationship is the same with and without SH0ES data, but with NEDE admitting a marginally larger value of $f_\mathrm{NEDE}$ than the corresponding value of $f_\mathrm{EDE}$ in the EDE model, as also found in Fig.~\ref{fig:f_comparison} and discussed above.

In terms of addressing the Hubble tension, the baseline constraints on $H_0$, given in Table~\ref{summary_table}, imply an alleviation of the Hubble tension from $\approx 4.8 \sigma$ to $\approx 2.1\sigma$ and $\approx 1.9\sigma$ in the NEDE and EDE models, respectively, using the Gaussian tension metric of Ref.~\cite{Schoneberg:2021qvd} (with the standard deviation approximated as the average of the upper and lower error bars). In terms of the $Q_\mathrm{DMAP}$ tension metric introduced in section~\ref{sec:baseline_sh0es}, the NEDE and EDE models also give alleviations from $\approx 4.8\sigma$ in $\Lambda$CDM to $\approx 2.1\sigma$ and $\approx 1.9 \sigma$, respectively (which we checked is approximately the same when computed from the $f_\mathrm{(N)EDE}$ and $H_0$ profiles). Thus, we find, unlike Ref.~\cite{Schoneberg:2021qvd}, that the Gaussian tension and the $Q_\mathrm{DMAP}$ metrics coincide: Our Gaussian metrics indicate a stronger alleviation than Ref.~\cite{Schoneberg:2021qvd} and vice-versa for the $Q_\mathrm{DMAP}$ estimates. To the extent that our baseline differs from that of the review in Ref.~\cite{Schoneberg:2021qvd}, the models retain their status as some of the most promising solutions to the Hubble tension.

Finally, we note that models of early dark energy may lead to increased values of $\sigma_8$ due to the correlation between $f_\mathrm{(N)EDE}$ and the cold dark matter density $\omega_\mathrm{cdm}$~\cite{Herold:2022iib,Niedermann:2020dwg}, thereby worsening the tension between the CMB inferences of $\sigma_8$ and low-redshift measurements, such as the Dark Energy Survey~\cite{DES:2021wwk}, of the same. We leave a profile likelihood analysis of $\sigma_8$ in the (N)EDE models for future work.

\subsection{Constraints on other parameters}\label{sec:constraints_on_other_parameters}
\begin{figure}[tb]
	\includegraphics[width=\columnwidth]{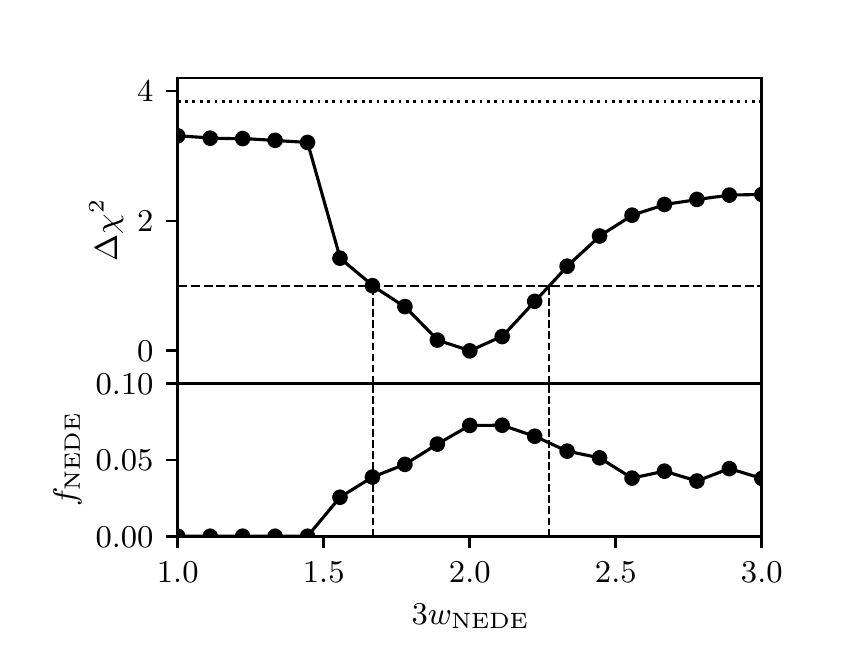}
	\caption{\label{fig:profile_3w} \textit{Top panel:} Profile likelihood for the NEDE equation of state $3w_\mathrm{NEDE}$. The horizontal dashed (dotted) lines corresponds to the value $\Delta \chi^2 = 1.0 (3.84)$, representing the $68\%$ $(95\%)$ confidence band. \textit{Bottom panel:} Values of the maximal NEDE fraction $f_\mathrm{NEDE}$ obtained by optimization at each point in the profile.}
\end{figure}
In addition to the above results, we have computed a profile likelihood in the equation of state of the NEDE field after its decay, $w_\mathrm{NEDE}$, under the baseline dataset. This is shown on Fig.~\ref{fig:profile_3w}, where the top panel shows the profile likelihood and the bottom panel shows the values of $f_\mathrm{NEDE}$ obtained from the optimization at each point in the profile. At $68 \%$ CL, we obtain the constraint $3 w_\mathrm{NEDE}=1.992^{+0.279}_{-0.323}$, i.e. an approximate bestfit equation of state of $w_\mathrm{NEDE} \approx 2/3$, corresponding to a relatively stiff fluid that redshifts faster than radiation. This is consistent with earlier findings~\cite{Niedermann:2020dwg, Cruz:2022oqk, Poulin:2021bjr}. Evidently, for smaller equations of state approaching the radiative limit of $1/3$, $\Lambda$CDM is entirely preferred over NEDE, as can be seen from the vanishing of $f_\mathrm{NEDE}$ at these points in the profile. On the other hand, there is a heavy tail towards larger values of the equation of state. The emerging picture is that the equation of state is relatively free as long as it is large enough for an adequately fast redshifting of the NEDE field after its decay. Note finally that the equation of state is entirely unconstrained at $95 \%$ CL, in agreement with fact that the value $f_\mathrm{NEDE}= 0$ lies inside the $95 \%$ CI as observed on Fig.~\ref{fig:intervals_profile}.

\begin{figure}[tb]
	\includegraphics[width=\columnwidth]{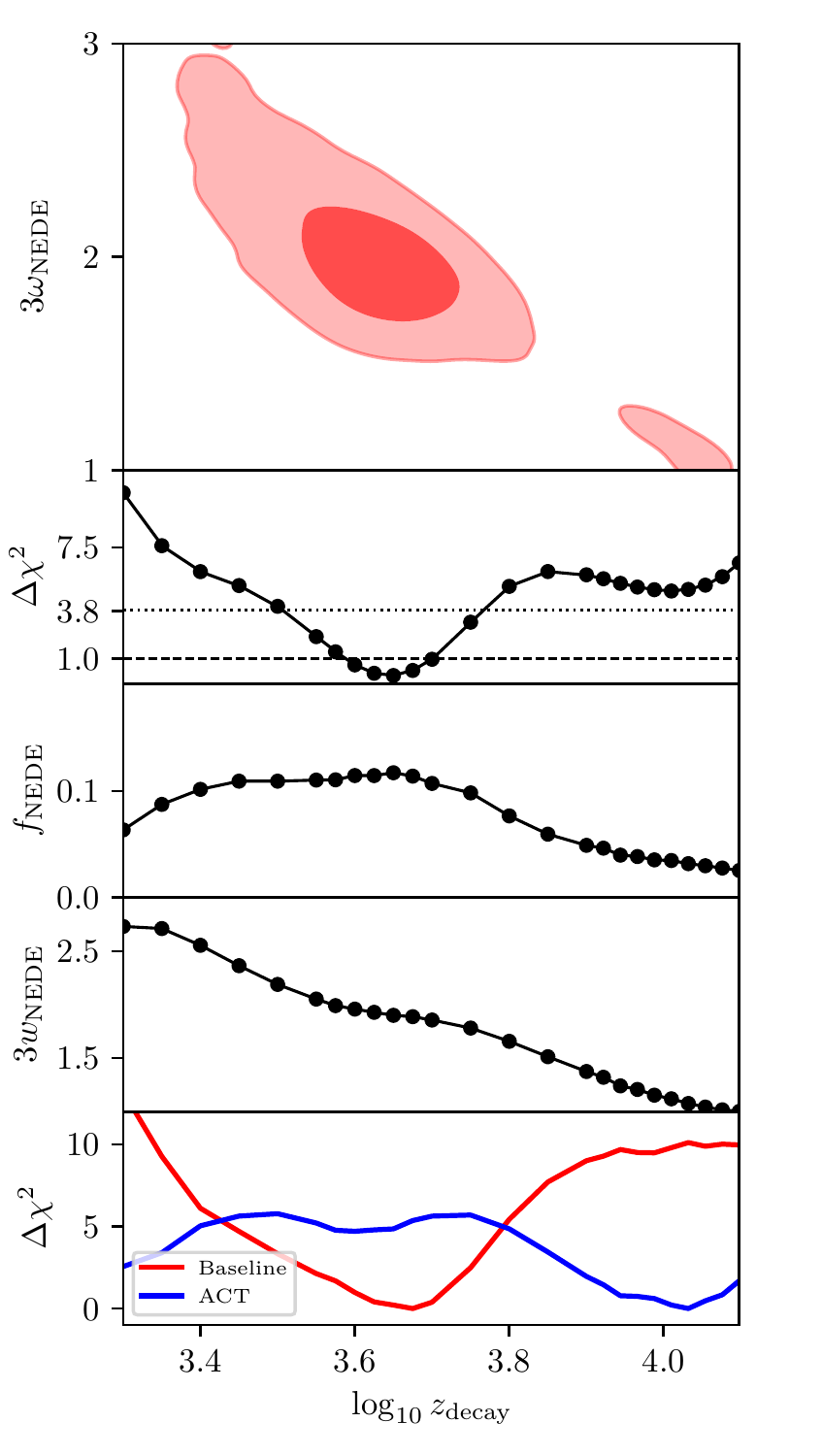}
	\caption{\label{fig:profile_log10z} Constraints on NEDE using the baseline, ACT and SH0ES data. \textit{Top panel:} $1\sigma$ and $2\sigma$ contours of the two-dimensional marginalised posterior of the NEDE parameters $3w_\mathrm{NEDE}$ and $\log_{10} z_\mathrm{decay}$ from Ref.~\cite{Cruz:2022oqk}. \textit{Second panel:} Profile likelihood in $\log_{10} z_\mathrm{decay}$. \textit{Third and fourth panels:} Values of $f_\mathrm{NEDE}$ and $3w_\mathrm{NEDE}$ obtained from optimization at each point in the profile of the second panel. \textit{Fifth panel:} Individual contribution of the baseline and ACT likelihoods to the $\chi^2$ budget at each point in the profile likelihood.}
\end{figure}
Ref.~\cite{Cruz:2022oqk} find a bimodality in the Bayesian posterior of NEDE when analysed with the baseline, ACT and SH0ES data. In addition to the main peak characterized by $w_\mathrm{NEDE} \approx 2/3$ and $\log_{10} z_\mathrm{decay} \approx 3.7$, the posteriors indicate a weaker mode (henceforth the \textit{bi-mode}) around $w_\mathrm{NEDE}\approx 1/3$ and $\log_{10} z_\mathrm{decay}\approx 4.0$ which is associated with a smaller value of $f_\mathrm{NEDE}$. To study this double peak structure, we have computed a one-dimensional profile likelihood in the decay time of NEDE, $\log_{10} z_\mathrm{decay}$, which is shown on Fig.~\ref{fig:profile_log10z}. The top panel of the figure shows the Bayesian two-dimensional marginalised posterior in the NEDE equation of state $3w_\mathrm{NEDE}$ and decay redshift $\log_{10} z_\mathrm{decay}$ obtained in Ref.~\cite{Cruz:2022oqk}. The second panel shows the profile likelihood in $\log_{10} z_\mathrm{decay}$, sampled with increased resolution around the two peaks. The third and fourth panels show the values of $f_\mathrm{NEDE}$ and $3w_\mathrm{NEDE}$ obtained through optimization at each point in the profile of the second panel. Lastly, the fifth panel shows the individual contribution to the $\chi^2$ budget of the likelihood from the baseline and ACT, respectively, at each point in the profile.

As seen on the second panel, the profile likelihood recovers the double peak structure of the Bayesian posterior. The dashed and dotted lines represent the values $\Delta \chi^2 = 1.0$ and $\Delta \chi^2 = 3.84$, respectively: As explained earlier, the intersection of these lines with the profile gives the approximate $68\%$ and $95\%$ confidence intervals. Apparently, the significance of the bi-mode is somewhat weaker in the profile than in the Bayesian analysis, hinting at a volume effect, but the difference is relatively small and may be due to computational uncertainties.

It is seen on the fifth panel of Fig.~\ref{fig:profile_log10z} that the the ACT likelihood exclusively prefers the bi-mode and that the baseline exclusively prefers the main mode. This tension between Planck and ACT was also noted in Ref.~\cite{Poulin:2021bjr}.


Although less significant, the bi-mode prefers intermediate values of $f_\mathrm{NEDE}$, possibly explaining our finding in section~\ref{sec:fNEDE_ACT} that smaller values of $f_\mathrm{NEDE}$ are more viable when including ACT. Since the bi-mode is exclusively driven by the ACT likelihood, fixing the trigger mass (or $z_\mathrm{decay}$, equivalently) as in Ref.~\cite{Cruz:2022oqk} to the value at the main peak diminishes the contribution of ACT to the total likelihood (since the ACT contribution is seen to be mostly flat around the main peak). Furthermore, since the ACT peak is associated with smaller values of $f_\mathrm{NEDE}$ than the main peak, this explains the fact that Ref.~\cite{Cruz:2022oqk} found a preference for larger values of $f_\mathrm{NEDE}$ than in our profile likelihood analysis.

\begin{table*}[tb]
	\centering
	\begin{tabular*}{0.7476\textwidth}{?Sc|Sc|l|Sc|Sc|Sc?}
		\specialrule{.12em}{0em}{0em}
		Data & Model & Parameter & $68 \%$ CL & $95 \%$ CL & Figure \\
		\hhline{|=|=|=|=|=|=|}
		\multirow{7}{*}{baseline} & \ \multirow{4}{*}{NEDE}  \hspace{1cm} & $f_\mathrm{NEDE}$ & \ $0.076^{+ 0.040}_{-0.035}$ \hspace{1cm} & \ $<0.154$ \hspace{1cm} & Fig.~\ref{fig:intervals_profile} \\ \cline{3-6}
		& & $H_0 \ [\mathrm{km} \ \mathrm{s}^{-1} \ \mathrm{Mpc}^{-1}]$ & $69.56^{+1.16}_{-1.29}$ & $69.56^{+2.41}_{-2.18}$ & Fig.~\ref{fig:profile_H0} \\ \cline{3-6}
		& & $3w_\mathrm{NEDE}$ & $1.992^{+0.279}_{-0.323}$ & unconstrained & Fig.~\ref{fig:profile_3w} \\ \cline{2-6}
		&  \multirow{2}{*}{EDE}  & $f_\mathrm{EDE}$ & $0.079^{+0.031}_{-0.040}$ & $<0.137$ & Fig.~\ref{fig:f_comparison} \\ \cline{3-6}
		& & $H_0 \ [\mathrm{km} \ \mathrm{s}^{-1} \ \mathrm{Mpc}^{-1}]$ & $70.02^{+1.20}_{-1.22}$ & $70.02^{+2.41}_{-2.38}$ & Fig.~\ref{fig:profile_H0} \\
		\hline
		\multirow{6}{*}{baseline + SH0ES} & \multirow{3}{*}{NEDE} & $f_\mathrm{NEDE}$ & $0.136^{+0.024}_{-0.026}$ & $0.136^{+0.047}_{-0.057}$ & Fig.~\ref{fig:intervals_profile} \\ \cline{3-6}
		& & $H_0 \ [\mathrm{km} \ \mathrm{s}^{-1} \ \mathrm{Mpc}^{-1}]$ & $71.62^{+0.78}_{-0.76}$ & $71.62^{+1.58}_{-1.55}$ & Fig.~\ref{fig:profile_H0} \\ \cline{2-6}
		& \multirow{2}{*}{EDE} & $f_\mathrm{EDE}$ & $0.112^{+0.030}_{-0.008}$ & $0.112^{+0.044}_{-0.030}$ & Fig.~\ref{fig:f_comparison} \\ \cline{3-6}
		& & $H_0 \ [\mathrm{km} \ \mathrm{s}^{-1} \ \mathrm{Mpc}^{-1}]$ & $71.80^{+0.72}_{-0.73}$ & $71.80^{+1.55}_{-1.55}$ & Fig.~\ref{fig:profile_H0} \\
		\hline
		\multirow{3}{*}{baseline + FS} & NEDE & $f_\mathrm{NEDE}$ & $0.084^{+0.034}_{-0.046}$ & $<0.157$ & Fig.~\ref{fig:intervals_profile} \\ \cline{2-6}
		& EDE & $f_\mathrm{EDE}$ & $0.069^{+0.030}_{-0.029}$ & $<0.137$ & Fig.~\ref{fig:lss} \\
		\hline
		baseline + ACT & NEDE & $f_\mathrm{NEDE}$ & $0.051^{+0.031}_{-0.030}$ & $<0.119$ & Fig.~\ref{fig:intervals_profile} \\
		\hline
		baseline + ACT + SH0ES & NEDE & $\log_{10} z_\mathrm{decay}$ & $3.645^{+0.056}_{-0.058}$ & $3.645^{+0.121}_{-0.139}$ & Fig.~\ref{fig:profile_log10z} \\
		\hline
		baseline + SPT & NEDE & $f_\mathrm{NEDE}$ & $0.061^{+0.038}_{-0.037}$ & $<0.136$ & Fig.~\ref{fig:intervals_profile} \\
		\specialrule{.12em}{0em}{0em}
	\end{tabular*}
	\caption{\label{summary_table} Summary of profile likelihood constraints obtained in this paper. All datasets are described in section~\ref{sec:level4}, and the confidence intervals are obtained from the Neyman method as described in section~\ref{sec:level3}.}
\end{table*}

\section{Conclusion}\label{sec:level5}
In this work, we have studied the NEDE extension of the $\Lambda$CDM model for the first time using profile likelihoods, which are inherently unaffected by the prior volume effects that tend to bias Bayesian constraints of $\Lambda$CDM extensions toward the $\Lambda$CDM limit~\cite{Holm:2022kkd, Herold:2022iib, Herold:2021ksg}. Our results are summarized in Table~\ref{summary_table}. The constraints on the maximal energy density fraction of NEDE, $f_\mathrm{NEDE}$, generally prefer larger values than the corresponding Bayesian analyses, corroborating the existence of volume effects in the model noted in Refs.~\cite{Niedermann:2020dwg,Niedermann:2020qbw,Cruz:2022oqk}. We have shown that when fixing the NEDE trigger field mass to its bestfit value, as done in the latter references to circumvent the volume effects, one obtains one-dimensional marginalised posterior distributions in $f_\mathrm{NEDE}$ that coincide with the profile likelihood. We therefore conclude that fixing the trigger field mass is an appropriate strategy to avoid the volume effects in NEDE. Using our baseline data, mainly dominated by \textit{Planck} CMB measurements, we find $f_\mathrm{NEDE}=0.076^{+0.040}_{-0.035}$ at $68 \%$ CL. Including full-shape power spectrum data from BOSS does not change the baseline constraints, indicating that \textit{Planck} remains the dominating constraint on NEDE. On the other hand, including high-$\ell$ CMB data from ACT gives similar constraints with a slightly smaller preference for NEDE due to a likelihood bimodality arising from ACT data.
 An analysis using the baseline and SPT data again provides a similar constraint on $f_\mathrm{NEDE}$.

Since the profile likelihood is reparametrization invariant, it is an excellent tool for comparing NEDE to the EDE-model. With our baseline, we find similar constraints on the maximal energy density fractions $f_\mathrm{NEDE}$ and $f_\mathrm{EDE}$, respectively, with NEDE admitting slightly larger values of $f_\mathrm{(N)EDE}$, possibly due to the fact that NEDE decays earlier than EDE around bestfit. We also find similar constraints on $H_0$ in the two models, with EDE being a slightly better fit at large values of $H_0$. Although EDE provides a marginally better global fit than NEDE, at $\chi^2_{\mathrm{min,NEDE}} - \chi^2_{\mathrm{min,EDE}} \approx -1.0$, both models reduce the Hubble tension from $\approx 4.8 \sigma$ to $\approx 2 \sigma$ and remain some of the most promising solutions to the Hubble tension~\cite{Schoneberg:2021qvd}.

\section*{Acknowledgements}
We thank Guido D'Amico for assistance in setting up \textsc{PyBird}. Additionally, we are very grateful to Vivian Poulin and Laura Herold for valuable discussions and comments on the draft. We acknowledge computing resources from the Centre for Scientific Computing Aarhus (CSCAA). E.B.H. and T.T. were supported by a research grant (29337) from VILLUM FONDEN. The work of F.N was supported by VR Starting Grant 2022-03160 of the Swedish Research Council. J.S.C and M.S.S. are supported by Independent Research Fund Denmark grant 0135-00378B.

\bibliographystyle{utcaps}
\bibliography{paper}

\end{document}